\documentclass[9pt,shortpaper,twoside,web]{ieeecolor}
\usepackage{generic}
\usepackage{cite}
\usepackage{amsmath,amssymb,amsfonts}
\usepackage{algorithmic}
\usepackage{graphicx}
\usepackage{textcomp}
\usepackage{xcolor}
\usepackage{color}
\usepackage{setspace}
\usepackage{subfigure}
\usepackage{caption}

\def\BibTeX{{\rm B\kern-.05em{\sc i\kern-.025em b}\kern-.08em
    T\kern-.1667em\lower.7ex\hbox{E}\kern-.125emX}}
\begin{document}
\title{Performance Analysis and ISI Mitigation with Imperfect Transmitter in Molecular Communication}
\author{Dongliang Jing, \IEEEmembership{Member, IEEE}, Lin Lin, \IEEEmembership{Senior Member,IEEE} and Andrew W. Eckford, \IEEEmembership{Senior Member,IEEE} 
\thanks{This work was supported by China Postdoctoral Science Foundation under Grant 2023M732877.}
\thanks{Dongliang Jing is with the College of Mechanical and Electronic Engineering, Northwest A\&F University, Yangling, China (e-mail: dljing@nwafu.edu.cn).}
\thanks{Lin Lin is with the College of Electronics and Information Engineering, Tongji University, Shanghai, China (e-mail: fxlinlin@tongji.edu.cn).}
\thanks{Andrew W. Eckford is with the Department of Electrical Engineering and Computer Science, York University, Toronto, Ontario, Canada (e-mail: aeckford@yorku.ca).}}

\maketitle

\begin{abstract}
In molecular communication (MC), molecules are released from the transmitter to convey information. This paper considers a realistic molecule shift keying (MoSK) scenario with two species of molecule in two reservoirs, where the molecules are harvested from the environment and placed into different reservoirs, which are purified by exchanging molecules between the reservoirs.
This process consumes energy, and for a reasonable energy cost, the reservoirs cannot be pure; thus, our MoSK transmitter is imperfect, releasing mixtures of both molecules for every symbol, resulting in inter-symbol interference (ISI). To mitigate ISI, the properties of the receiver are analyzed and a detection method based on the ratio of different molecules is proposed. Theoretical and simulation results are provided, 
showing that with the increase of energy cost, the system achieves better performance. The good performance of the proposed detection scheme is also demonstrated.
\end{abstract}

\begin{IEEEkeywords}
Molecular communication (MC), imperfect transmitter, energy cost, inter symbol interference (ISI).
\end{IEEEkeywords}

\section{Introduction}
Molecular communication (MC), which is inspired by biological communication, employs molecules as information carriers between nanomachines \cite{nakano2013molecular,farsad2016comprehensive,jamali2019channel}. MC is envisioned as a promising approach to the problem of communication in nanoscale networks \cite{kim2014symbol}. MC has
 various potential healthcare applications in health monitoring, disease diagnosis, and targeted drug delivery\cite{bao2021relative}.
 MC has also been employed to model the spread of infectious
disease via aerosols during the outbreak of COVID-19 \cite{khalid2020modeling,schurwanz2021infectious,chen2022detection}. 

In MC, as the molecules released from the transmitter diffuse freely to the receiver without external energy, diffusion-based molecular communication (DMC) is highly energy efficient in terms of propagation. However, due to the small size of the nanomachine, its energy reserves are restricted. Therefore, energy efficiency becomes a significant constraint in DMC, so energy cost in DMC can not be ignored, especially in creating the transmitter. {\color{black}Related works} in the literature have considered transmitter {\color{black}design} in terms of molecule synthesis: for example, an energy model for DMC considering the energy expended on molecule synthesis, secretory vesicle production, and vesicle transportation to the cell membrane is proposed in \cite{kuran2010energy}. In \cite{pierobon2012capacity},  thermodynamic energy consumption associated with particle emission in DMC signal transmission is studied. In \cite{eckford2018thermodynamic}, from the thermodynamic properties, the energy cost of DMC is studied without considering the diffusion of the molecules. To improve the energy efficiency, in \cite{tepekule2015isi}, a power adjustment technique and a decision feedback filter are proposed. In \cite{musa2020lean}, a DMC powered by a nanoscale energy-harvesting mechanism is studied.

In MC, the transmitter can either generate signaling molecules as needed or have a reservoir of such molecules available for transmission \cite{farsad2016comprehensive,kuscu2019transmitter}. 
In this paper, we focus on the latter case, where signaling molecules are stored in a reservoir, and harvested from the environment \cite{bilgin2017fast}. Even if molecules are synthesized as needed \cite{kuran2010energy,cheng2020energy}, for example by genetically engineered bacteria \cite{unluturk2015genetically}, it is still necessary to store them at least briefly. 
Furthermore, the molecules in these reservoirs are not necessarily pure: in addition to the intended molecules, there also exist interfering molecules. In biological systems, it is common for cells or tissues to release information molecules that also contain other types of molecules.
A specific example of such a phenomenon can be observed with Antimicrobial Peptides (AMPs). AMPs are a class of small molecular proteins that exhibit antimicrobial activity and contribute to immune defense \cite{nordstrom2017delivery,teixeira2020nanomedicines}. In certain cases, a single cell has the capability to release two types of AMPs. Some AMPs act as signaling molecules, influencing the behavior of other cells and modulating immune responses. On the other hand, there are AMPs that function as interfering molecules, targeting and disrupting the physiological processes of pathogens. This interference can occur through various mechanisms such as the disruption of pathogen cell membranes, interference with their metabolism, or inhibition of their growth \cite{li2021plant}.

In MC, information is encoded in the physical properties of the molecules (e.g. the number of molecules, the type of molecules, or the release time of the molecules). In this paper we use the type of molecules to encode information, a method known as molecule shift keying (MoSK). For example, in binary molecular shift keying (BMoSK), two types of molecules are required to encode the information: bit 0 is expressed with molecule A, while bit 1 is expressed with molecule B; the receiver can then decide which bit was sent based on whether it observes more of one molecule than the other.
Variants of MoSK are actively studied. A depleted
MoSK scheme is proposed in \cite{kabir2015d}, where multiple types of molecules are simultaneously employed to transmit the parallel streams of concentration shift keying for each symbol.
In \cite{7331300}, a molecular transition shift keying scheme is proposed, in which two types of molecules can be released, but the released molecule type is chosen based on previous transmissions.
In \cite{shi2020performance}, the performance of DMC systems with BMoSK modulation was analyzed.
In \cite{wen2021layered}, a layered molecular shift keying (LMoSK) modulation scheme was proposed in which each type of molecule was used to define a communication layer. 
{\color{black} }

In MC, diffusion results in channel memory and inter-symbol interference (ISI), which reduce the information rate \cite{tepekule2015isi}. To mitigate these issues, modulation/channel coding schemes \cite{7784766,kislal2019isi,keshavarz2019inter,tang2020molecular,gursoy2021concentration} and detection schemes \cite{li2016local,chang2017adaptive,noel2014optimal,singhal2015performance,9887933,ghavami2017abnormality,fang2019symbol,qian2021k} have been proposed. Arjmandi et al. \cite{7784766} proposed an ISI-avoiding modulation scheme by enlarging the time instances of the same type of molecules.
In \cite{tang2020molecular}, molecular type permutation shift keying was proposed, which mitigates ISI by encoding the information bits in permutations of diverse types of molecules.
A precoder scheme based on the concentration difference of two types of molecules was also proposed to mitigate the ISI. 
In \cite{gursoy2021concentration}, a hybrid modulation scheme was proposed to mitigate ISI based on molecules pulse position and concentration. In \cite{aghababaiyan2022enhanced}, a binary direction shift keying modulation scheme is proposed to mitigate multiuser interference, wherein molecules are released in two different directions.
By studying the local convexity of the received signals,  Li et al. \cite{li2016local} proposed to detect the received symbols based on molecular concentration difference. Chang et al. \cite{chang2017adaptive} proposed an adaptive detection scheme to mitigate ISI based on the reconstruction of the channel impulse response.
Maximum likelihood (ML) detection schemes \cite{noel2014optimal,singhal2015performance,9887933} and symbol-by-symbol ML detection schemes \cite{ghavami2017abnormality,fang2019symbol} have also been proposed in molecular communication.  Moreover,
non-coherent detection schemes based on K-means clustering were studied in \cite{qian2021k}. Further machine learning methods have been proposed to mitigate the ISI in molecular communication \cite{qian2018receiver,qian2019molecular}. However, for machine learning schemes, it is challenging to obtain substantial amounts of data for training.

In this paper, based on the works from \cite{eckford2018thermodynamic}, {\color{black}which investigated the purification process involving the movement of molecules between reservoirs, it was revealed that the establishment of reservoirs incurs a free energy cost. However, due to practical energy constraints, achieving complete separation of different molecule types becomes infeasible; as a result, it is necessary to tolerate partially impure reservoirs. This impurity, in turn, causes significant interference at the receiver. Consequently, we address an MC system with an imperfect transmitter featuring two reservoirs. Within these reservoirs, molecules are mixed, and distinctions between the two molecule types are created by moving molecules between the reservoirs. This transfer process necessitates energy input, and due to practical energy considerations, the reservoirs contain interfering molecules, which leads to severe interference at the receiver.} Unlike \cite{huang2021membrane} where the errors in an imperfect transmitter arise from the molecules being probabilistically released from the transmitter membrane, and unlike \cite{eckford2018thermodynamic}, which primarily centers on the energy consumption associated with moving molecules between reservoirs, this paper provides an analysis of system performance in the presence of an imperfect transmitter, ISI, and counting noise. Additionally, a detection method is proposed to mitigate ISI based on the ratio of different molecules.

The main contributions of this paper are summarized as follows.
\begin{itemize}
\item An imperfect transmitter is considered, where one kind of molecule is moved from one reservoir to the other and the energy consumption is analyzed.   
\item As the released molecules contain interfering molecules, we analyze the properties of the received molecules under the imperfect transmitter and the channel memory.
\item A detection method based on the ratio of different types of molecules is proposed to mitigate the ISI. 
\end{itemize}

The remainder of this paper is organized as follows. In section II, we introduce the system model of the considered imperfect transmitter MC system. In section III, the performance of the proposed imperfect transmitter MC system is analyzed. Numerical and simulation results are presented in Section IV. Finally, in Section V, we conclude this paper.
\section{System Model}

Consider a system, analogous to the one described in \cite{eckford2018thermodynamic}, wherein the environment comprises a mixture of two molecular types, denoted as $\rm{A}$ and $\rm{B}$. In this context, we make the assumption that $\rm{A}$ and $\rm{B}$ exclusively constitute the molecular composition of the environment. The initial concentration of these molecules is quantified by a molar fraction of $\rm{B}$, denoted as $c$. {\color{black}This molar fraction, a metric expressing the concentration of $\rm{B}$ relative to the overall concentration of both $\rm{A}$ and $\rm{B}$, is defined as the ratio of the moles of $\rm{B}$ to the total moles of both $\rm{A}$ and $\rm{B}$ molecules present in the environment.}

In the binary MoSK molecular communication system, two molecular reservoirs are required to store two types of molecules. In this paper, we assume the information molecules collected from the environment are mixed, i.e., each reservoir can contain both kinds of molecules, though at different concentrations.
Initially, the contents of the reservoirs are acquired from the environment, resulting in equal concentrations of each type of molecule in the low and high reservoirs. In order to use MoSK, we need a different concentration in each reservoir, so that one primarily contains one molecule and the other primarily contains the other, although in this scenario the concentrations will be impure. Thus,
we propose moving one molecular type from one reservoir to the other. This creates a concentration difference between the two reservoirs, allowing the detector to decode information based on the ratio of the two molecular concentrations. However, it is important to note that the act of purifying the reservoirs requires energy. This process increases the chemical potential of the reservoirs with respect to the environment and requires an investment of free energy \cite{leff2002maxwell}. Thus, even though no energy is required to synthesize molecules, the thermodynamic properties of the system require a fundamental amount of energy in order to purify the reservoirs. 

\begin{figure}[!t]
  \centering
  \includegraphics[width=0.47\textwidth]{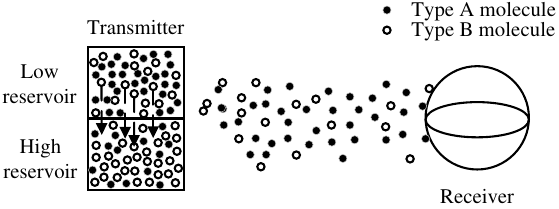}\\
  \caption{A depiction of the communication system. $\rm{A}$ molecules are filled circles, while $\rm{B}$ molecules are unfilled circles (cf. \cite[Fig. 2]{eckford2018thermodynamic}). Both the low reservoir and high reservoir are filled with $\rm{A}$ and $\rm{B}$ molecules. {\color{black}By moving B molecules from the low reservoir to the high reservoir, a difference in the concentrations of A and B molecules is created between the two reservoirs.}} \label{communication system}
\end{figure}

The transmitter creates two reservoirs of molecules from the environmental mixture, both initially at concentration $c$. 
From these, it creates a {\em difference of concentration} between the two reservoirs by transferring molecules of $\rm{B}$ from one to the other: the donating reservoir is called the {\em low reservoir}, with concentration $c_L$; while the receiving reservoir is called the {\em high reservoir}, with concentration $c_H$. (The terms {\em low} and {\em high} refer to the concentration of $\rm{B}$, which is low in the donating reservoir and high in the receiving reservoir.) {\color{black}However, due to practical energy constraints during the moving process, complete separation of different molecule types becomes infeasible. This results in reservoirs containing interfering molecules that are subsequently released, contributing to the occurrence of severe ISI at the receiver.} The transmitter sends a $\{0,1\}$ bit by selecting molecules from either the low reservoir (to send 0) or the high reservoir (to send 1). The communication system is depicted in Figure \ref{communication system}. The reservoirs are replenished at the beginning of the communication session and contain a sufficient number of molecules in order to transmit all required symbols. {\color{black} To create the concentration difference of A and B molecules between the low and high reservoirs, B molecules are moved from the low reservoir to the high reservoir, a process which consumes free energy.} The ratio in each reservoir is not actively controlled, but the method of selecting molecules for transmission will maintain the concentration constant, on average, in each reservoir.

Creating the reservoir requires an input of free energy because of the changing chemical potential. From \cite{eckford2018thermodynamic}, if $m$ molecules of type $\rm{B}$ are moved from the low reservoir to the high reservoir (where $m$ is small compared to the total number of molecules in the reservoir), then the energy cost can be expressed as
\cite[Eqn. (25)]{eckford2018thermodynamic} 
\begin{equation}
\label{equ3}
E = {n_H}kT_e\left[ { - \left( {1 + \frac{{{n_L}}}{{{n_H}}}} \right)c\log c + {c_H}\log {c_H} + \frac{{{n_L}}}{{{n_H}}}{c_L}\log {c_L}} \right],   
\end{equation} 
where $n_L$ and $n_H$ are the total number of molecules in the low and high reservoirs, respectively,  $k$ is Boltzmann's constant, $T_e$ is the absolute temperature, $c_L$ and $c_H$ are the fraction of $\rm{B}$ molecules in the low and high reservoirs, respectively. Then, {\color{black}after moving $m$ molecules of type $\rm{B}$ from the low reservoir to the high reservoir,} the mole fractions of type $\rm{B}$ molecules in the low and high reservoirs can be {\color{black}approximated as}
\begin{align}
\label{equ2}
\begin{split}
{c_L} = c - \frac{m}{{{n_L}}},\\
{c_H} = c + \frac{m}{{{n_H}}}.
\end{split}
\end{align}

To analyze the performance of our transmission scheme, from (\ref{equ3}), 
we focus on how the number of moved molecules $m$ varies with the energy $E$.
Assume that $n_L=n_H$, and let $n = n_L+n_H$ represent the total number of molecules in both reservoirs. Then (\ref{equ3}) can be expressed as
\begin{align}
\label{equ4}
\begin{split}
E  = &{n_H}kT_e\left[ \left( {c + \frac{{2m}}{n}} \right)\log \left( 1 + \frac{{2m}}{{cn}}\right) \right. \\ 
&+ \left. \left( c - \frac{{2m}}{n} \right)\log \left( 1 - \frac{{2m}}{{cn}} \right) \right].
\end{split}  
\end{align}

To simplify the notation, define $\alpha = \frac{{2m}}{n}$, the number of molecules moved as a fraction of the reservoir size, and define
$\beta  = \frac{\alpha }{c} = \frac{{2m}}{{cn}}$. Then
(\ref{equ4}) can be expressed as
\begin{align}
\label{equ5}
E = {n_H}kT_e\left[ {\left( {c + \alpha } \right)\log \left( {1 + \beta } \right) + \left( {c - \alpha } \right)\log \left( {1 - \beta } \right)} \right].    
\end{align}
By assumption, the number of moved molecules $m$ is much smaller than the number of molecules in the low or high reservoir. Thus, $0 < \beta \ll 1$. By employing a Taylor series \cite{5273711,5264196}, we can write
\begin{align}
\label{equ6}
\begin{split}
\log \left( {1 + \beta } \right) = \beta  - \frac{1}{2}{\beta ^2} + o\left( {{\beta ^3}} \right), \\
\log \left( {1 - \beta } \right) =  - \beta  - \frac{1}{2}{\beta ^2} + o\left( {{\beta ^3}} \right).
\end{split}
\end{align}
Therefore, (\ref{equ5}) can be expressed as
\begin{align}
\label{equ7}
\begin{split}
E &= {n_H}kT_e\left[ {\left( {c + \alpha } \right)\left( {\beta  - \frac{1}{2}{\beta ^2}} \right) + \left( {c - \alpha } \right)\left( { - \beta  - \frac{1}{2}{\beta ^2}} \right)} \right] \\
&+ o\left( {{\beta ^3}} \right)\\ 
 &= {n_H}kT_e\left[ { - c{\beta ^2} + 2\alpha \beta } \right] + o\left( {{\beta ^3}} \right)\\ 
 &= {n_H}kT_e\left[ { - c{{\left( {\frac{{2m}}{{cn}}} \right)}^2} + 2\frac{{2m}}{n}\left( {\frac{{2m}}{{cn}}} \right)} \right] + o\left( {{\beta ^3}} \right)\\
 &= 2kT_e\frac{{{m^2}}}{{c{n}}} + o\left( {{\beta ^3}} \right).
\end{split}
\end{align}
From now on we will ignore the $o\left( {{\beta ^3}} \right)$ term.
Therefore, for the given energy cost, the number of moved molecules $m$ can be expressed as
\begin{align}
\label{equ8}
m = \sqrt {\frac{{cn}}{{2kT_e}}E}.
\end{align}
Then, based on (\ref{equ2}), $c_L$ and $c_H$ can be expressed as
\begin{align}
\begin{split}
{c_L} = c - \sqrt {\frac{{cn}}{{2kT_e{n_L}}}E}, \\
{c_H} = c + \sqrt {\frac{{cn}}{{2kT_e{n_H}}}E}.   
\end{split}    
\end{align}
The number of ${\rm{A}}$ and ${\rm{B}}$ molecules in the low and high reservoirs are respectively
\begin{align}
\begin{split}
{N_{L,{\rm{A}}}} = (n_L-m)(1-c_L),\\
{N_{L,{\rm{B}}}} = (n_L-m)c_L,\\
{N_{H,{\rm{A}}}} = (n_H+m)(1-c_H),\\
{N_{H,{\rm{B}}}} = (n_H+m)c_H.\\
\end{split}
\end{align}

After moving $m$ molecules of type $\rm{B}$ from the low reservoir to the high reservoir, 
there is a concentration difference of $\rm{B}$ molecules between the low and high reservoir, and can be expressed as 
$ {c_{L,{\rm{A}}}}/{c_{L,{\rm{B}}}} > {c_{H,{\rm{A}}}}/{c_{H,{\rm{B}}}}$, 
where ${c_{L,{\rm{A}}}}$ and ${c_{L,{\rm{B}}}}$ are the concentration of ${\rm{A}}$ and ${\rm{B}}$ molecules in the low reservoir,  ${c_{H,{\rm{A}}}}$ and ${c_{H,{\rm{B}}}}$ are the concentration of ${\rm{A}}$ and ${\rm{B}}$ molecules in the high reservoir, respectively. 
In the communication process, MoSK is employed. Therefore, during the $k$th bit interval, to transmit $N_{tx}$ molecules from the low reservoir for bit $0$, the transmitted signal can be expressed as
\begin{align}
x\left(t\right) = N_{tx,k}\varpi_{L,{\rm{A}}}+N_{tx,k}\varpi_{L,{\rm{B}}},  
\end{align}
where $\varpi$ is a random variable representing the fraction of selected molecules of each type, i.e., $\varpi_{L,{\rm{A}}}$ is the fraction of $\rm{A}$ molecules in the selected molecules (where the mean fraction of $\rm{A}$ molecules in the low reservoir is ${1 - {c_L}}$, which is the transmitted signal); and $\varpi_{L,{\rm{B}}}$ is the fraction of $\rm{B}$ molecules in the selected molecules (where
the mean fraction of $\rm{B}$ molecules in the low reservoir is ${c_L}$, which is interference). 

Similarly, to transmit $N_{tx,k}$ molecules from the high reservoir for bit $1$, the transmitted signal can be expressed:
\begin{align}
x\left(t\right) = N_{tx,k}\varpi_{H,{\rm{A}}}+N_{tx,k}\varpi_{H,{\rm{B}}},
\end{align}
where $\varpi_{H,{\rm{A}}}$ is the fraction of $\rm{A}$ molecules in the selected molecules and
the mean fraction of $\rm{A}$ molecules in the high reservoir is ${1 - {c_H}}$, which is interference; $\varpi_{H,{\rm{B}}}$ is the fraction of $\rm{B}$ molecules in the selected molecules and
the mean fraction of $\rm{B}$ molecules in the high reservoir is ${c_H}$, which is intend to transmit.

\section{Performance analysis}

Our goal in this paper is to find a tradeoff between the chemical potential energy required to create the transmitter, and the bit error rate from using the transmitter; thus, in this section, we derive an expression for the probability of error. Compared to the traditional MoSK, the performance analysis in the considered MC system with an imperfect transmitter is more challenging, as the transmitted molecules are mixed with $\rm{A}$ and $\rm{B}$ molecules whether for bit 0 and bit 1.  Meanwhile, as the released molecules include the intended molecules and the other type of molecules, which are interference molecules, therefore, the released mixed molecules make more severe ISI compared to the traditional MoSK in MC, resulting in the BER performance decrease. In this section, we focus on the analysis of the system performance of the considered MC system with the imperfect transmitter. Assuming the probability of transmitting bit 0 is $\epsilon$, so the probability of transmitting bit 1 is $1-\epsilon$. 
\subsection{Analysis of the received molecules}
Compared to the distance between the transmitter and the receiver, the size of the information molecules and reservoirs are assumed to be very small, then, the transmitter is modeled by a point transmitter. Considering a 3D absorbing receiver, the fraction of molecules absorbed by the receiver until time $t$, $F_{\rm{hit}}\left(t\right)$, can be expressed as
\begin{align}
{F_{\rm{hit}}}\left( t \right) = \frac{r}{{d + r}}\text{erfc}\left( {\frac{d}{{\sqrt {4D_mt} }}} \right),  
\end{align}
where $r$ is the radius of the receiver, $d$ is the distance between the transmitter and the receiver, $D_m$ is the diffusion coefficient of the released molecules (The diffusion coefficient of $\rm{A}$ and $\rm{B}$ molecules are assumed to be the same, which is a common assumption in MC, such as for isomers \cite{kim2013novel}). Considering $N_{tx,k}$ molecules are released by the transmitter and the transmitted signal, the the average number of received molecules can be expressed as
\begin{align}
y\left(t\right) = x\left(t\right) F_{\rm{hit}}\left( t \right) .  
\end{align}

During the $k$th bit interval, the hitting probability $q_k$ can be expressed as
\begin{align}
q_k = F_{\rm{hit}}\left(kt_s\right)-F_{\rm{hit}}\left(\left[k-1\right]t_s\right), \end{align}
where $t_s$ is the bit interval. 

%
%

As the transmission contains $\rm{A}$ and $\rm{B}$ molecules, the mean number of received $\rm{A}$ molecules can be expressed as
\begin{align}
\label{received_molecules_A}
{N_{rx,k,{\rm{A}}}} = {N_{c,k,{\rm{A}}}} + {N_{ \text{ISI},k,{\rm{A}}}} + {N_{n,k,{\rm{A}}}},  
\end{align}
where $N_{c,k,{\rm{A}}}$ is the number of received $\rm{A}$ molecules during the $k$th bit interval which molecules transmitted at the beginning of $k$th bit interval, $N_{ \text{ISI},k,{\rm{A}}}$ is the number of $\rm{A}$ molecules received during the $k$th bit interval but transmitted from the previous bit interval and defined as ISI, $N_{n,k,{\rm{A}}}$ is the counting noise of $\rm{A}$ molecules which can be modeled by the Gaussian distribution ${N_{n,k,{\rm{A}}}} \sim {\mathcal N}\left( {0,\sigma _{n,k,{\rm{A}}}^2} \right)$, where $\sigma _{n,k,{\rm{A}}}^2$ depends on the average number of received $\rm{A}$ molecules. The quantity $N_{c,k,{\rm{A}}}$ can be expressed as 
\begin{align}
N_{c,k,{\rm{A}}}= {N_{tx,k}}{c_{tx,k,{\rm{A}}}}{q_1},  
\end{align}
where ${N_{tx,k}}$ is the transmitted molecules in the $k$th bit interval and selected from the low or high reservoir; ${c_{tx,k,{\rm{A}}}}$ is the fraction of $\rm{A}$ molecules in the selected reservoir; 
$q_1$ is the probability of molecules transmitted at the beginning of the $k$th bit interval and observed by the receiver during the $k$th bit interval.

The number of molecules received as ISI, $N_{\text{ISI},k,{\rm{A}}}$, can be expressed as 
\begin{align}
\label{received_ISI_A}
{N_{\text{ISI},k,{\rm{A}}}} = \sum\limits_{i = 1}^{k - 1} {{I_{i,{\rm{A}}}}},  
\end{align}
where $I_{i,{\rm{A}}}$ is the number of ISI molecules which are transmitted at the $i$th bit interval and can be expressed as
\begin{align}
I_{i,{\rm{A}}} =  N_{tx,i}{c_{tx,i,{\rm{A}}}}q_{k-i+1},
\end{align}
where $q_{k-i+1}$ is the probability of molecules transmitted at the beginning of $i$th bit interval and observed by the receiver during the $k$th bit interval. As the number of transmitted molecules is large, then, $I_{i,{\rm{A}}}$ can be approximated by a normal distribution: ${I_{i,\rm{A}}} \sim {\mathcal N}\left( N_{tx,i}{c_{tx,i,{\rm{A}}}}q_{k-i+1}, N_{tx,i}{c_{tx,i,{\rm{A}}}}q_{k-i+1}\left(1-q_{k-i+1}\right) \right)$. Therefore, $N_{\text{ISI},k,{\rm{A}}}$ can also be approximate by a normal distribution.

Similarly, the mean number of received $\rm{B}$ molecules can be expressed as 
\begin{align}
{N_{rx,k,{\rm{B}}}} = {N_{c,k,{\rm{B}}}} + {N_{ \text{ISI},k,{\rm{B}}}} + {N_{n,k,{\rm{B}}}},  
\end{align}
where
\begin{align}
N_{c,k,{\rm{B}}} = N_{tx,k}{c_{tx,k,{\rm{B}}}}q_1,  
\end{align}
\begin{align}
{N_{\text{ISI},k,{\rm{B}}}} = \sum\limits_{i = 1}^{k - 1} {{I_{i,{\rm{B}}}}},  
\end{align}
and $I_{i,{\rm{B}}}$ can be expressed as
\begin{align}
I_{i,{\rm{B}}} &=  N_{tx,i}{c_{tx,i,{\rm{B}}}}q_{k-i+1}.
\end{align}
Similarly, $I_{i,{\rm{B}}}$ can be approximated by a normal distribution: ${I_{i,\rm{B}}} \sim {\mathcal N}\left( N_{tx,i}{c_{tx,i,{\rm{B}}}}q_{k-i+1}, N_{tx,i}{c_{tx,i,{\rm{B}}}}q_{k-i+1}\left(1-q_{k-i+1}\right) \right)$. Therefore, $N_{\text{ISI},k,{\rm{B}}}$ can also be approximate by a normal distribution. $N_{n,k,{\rm{B}}}$ is the counting noise of $\rm{B}$ molecules which can be modeled by the Gaussian distribution ${N_{n,k,{\rm{B}}}} \sim {\mathcal N}\left( {0,\sigma _{n,k,{\rm{B}}}^2} \right)$, where $\sigma _{n,k,{\rm{B}}}^2$ depends on the average number of received $\rm{B}$ molecules.




%
%
%
\subsection{Detection and ISI mitigation}
The detector can be formulated by the binary hypothesis testing problem with the received ${\rm{A}}$ and ${\rm{B}}$ molecules. Though in \cite{jing2022extended}, a binary hypothesis testing scheme has been employed, however, different to \cite{jing2022extended}, in this paper, due to the imperfect transmitter, the transmitted molecules are mixed, making the analysis of received molecules more complex.
The hypotheses are given by:
\begin{itemize}
\item ${\mathcal{H}_0}$: molecules are emitted from the low reservoir. $N_{rx,k,{\rm{A}}} \sim f_{N_{rx},{\rm{A}}}^{\mathcal{H}_0}$;$N_{rx,k,{\rm{B}}} \sim f_{N_{rx},{\rm{B}}}^{\mathcal{H}_0}$,
\item ${\mathcal{H}_1}$: molecules are emitted from the high reservoir;
$N_{rx,k,{\rm{A}}} \sim f_{N_{rx},{\rm{A}}}^{\mathcal{H}_1}$;$N_{rx,k,{\rm{B}}} \sim f_{N_{rx},{\rm{B}}}^{\mathcal{H}_1}$;
\end{itemize}
where $f_{N_{rx},{\rm{A}}}^{\mathcal{H}_0}$ and $f_{N_{rx},{\rm{B}}}^{\mathcal{H}_0}$ denote the probability density function of received $\rm{A}$ and $\rm{B}$ molecules under ${\mathcal{H}_0}$, respectively, while $f_{N_{rx},{\rm{A}}}^{\mathcal{H}_1}$ and $f_{N_{rx},{\rm{B}}}^{\mathcal{H}_1}$ denote the probability density function of received $\rm{A}$ and $\rm{B}$ molecules under ${\mathcal{H}_1}$, respectively.  As the $\rm{A}$ and $\rm{B}$ molecules select from the low or high reservoir by probability and the number of transmitted molecules is large, therefore, $N_{rx,k,{\rm{A}}}$ and $N_{rx,k,{\rm{B}}}$  can be approximated by the normal distribution \cite{kuran2010energy}. 

Under hypothesis $\mathcal{H}_0$, the mean number of received $\rm{A}$ and $\rm{B}$ molecules can be expressed as
\begin{equation}
\begin{split}
\label{received_A}
{N_{rx,k,{\rm{A}}}} =& {N_{tx,k}}\left( {1 - {c_L}} \right){q_1}\\ 
&+ \sum\limits_{i = 1}^{k - 1} {{N_{tx,i}}{c_{tx,i,{\rm{A}}}}{q_{k-i + 1}} + {N_{n,k,{\rm{A}}}}}, \\
\end{split}     
\end{equation}
and
\begin{equation}
\begin{split}
{N_{rx,k,{\rm{B}}}} =& {N_{tx,k}}{c_L}{q_1} \\
&+ \sum\limits_{i = 1}^{k - 1} {{N_{tx,i}}{c_{tx,i,{\rm{B}}}}{q_{k-i + 1}} + {N_{n,k,{\rm{B}}}}},
\end{split}     
\end{equation}
where ${c_{tx,i,{\rm{A}}}}$ and ${c_{tx,i,{\rm{B}}}}$ are the fraction of $\rm{A}$ and $\rm{B}$ molecules in the selected reservoir, respectively.
And the number of received $\rm{A}$ and $\rm{B}$ molecules under $\mathcal{H}_0$ can be approximated by the normal distribution,
${N_{rx,k,{\rm{A}}}} \sim \mathcal{N}\left( {{\mu _{{\mathcal{H}_0},{\rm{A}}}},\sigma _{{\mathcal{H}_0},{\rm{A}}}^2} \right)$, ${N_{rx,k,{\rm{B}}}} \sim \mathcal{N}\left( {{\mu _{{\mathcal{H}_0},{\rm{B}}}},\sigma _{{\mathcal{H}_0},{\rm{B}}}^2} \right)$. The mean and variance of received $\rm{A}$ molecules under $\mathcal{H}_0$ during the $k$th bit interval can be expressed as
\begin{align}
\begin{split}
\label{mu_H0_A}
{\mu _{{\mathcal{H}_0},k,{\rm{A}}}} =& {N_{tx,k}}\left( {1 - {c_L}} \right){q_1} + \sum\limits_{i = 1}^{k - 1} \left[ \epsilon {N_{tx,i}}\left( {1 - {c_L}} \right){q_{k-i + 1}} \right.\\
&\left. + \left( {1 - \epsilon } \right){N_{tx,i}}\left( {1 - {c_H}} \right){q_{k-i + 1}} \right],
\end{split}    
\end{align}
and 
as the symbols are selected independently and propagation of each molecule is independent, the event that a molecule arrives at the receiver in a particular interval is independent of every other molecule, thus, the sum of these events as in equation ( \ref{mu_H0_A}) is also independent, so the variance of the sum is the sum of the underlying variances.
\begin{align}
\begin{split}
\label{sigma_H0_A}
\sigma _{{\mathcal{H}_0},k,{\rm{A}}}^2 = {N_{tx,k}}\left( {1 - {c_L}} \right){q_1}\left( {1 - {q_1}} \right) + \sum\limits_{i = 1}^{k - 1} {\sigma _{{\rm{ISI}},i,{\rm{A}}}^2}  + \sigma _{n,k,{\rm{A}}}^2,   
\end{split}    
\end{align}
where
\begin{align}
\begin{split}
\label{sigma_ISI_A}
\sigma _{{\rm{ISI}},i,{\rm{A}}}^2 
 =& \epsilon \left( {1 - \epsilon } \right){\left[ {{N_{tx, i}}\left( {1 - {c_L}} \right){q_{k-i + 1}}} \right]^2} \\
 &+ \epsilon \left( {1 - \epsilon } \right){\left[ {{N_{tx, i}}\left( {1 - {c_H}} \right){q_{k-i + 1}}} \right]^2}\\
 &+ \epsilon {N_{tx,i}}\left( {1 - {c_L}} \right){q_{k-i + 1}}\left( {1 - {q_{k-i + 1}}} \right) \\
 &+ \left( {1 - \epsilon } \right){N_{tx,i}}\left( {1 - {c_H}} \right){q_{k-i + 1}}\left( {1 - {q_{k-i + 1}}} \right)\\
 &- 2\epsilon \left( {1 - \epsilon } \right){N_{tx,i}}\left( {1 - {c_L}} \right){q_{k-i + 1}}\\
 &\times {N_{tx,i}}\left( {1 - {c_H}} \right){q_{k-i + 1}}.
\end{split}    
\end{align}
The mean and variance of received $\rm{B}$ molecules under $\mathcal{H}_0$ during the $k$th bit interval can be expressed as 
\begin{align}
\begin{split}
\label{mu_H0_B}
{\mu _{{\mathcal{H}_0},k,{\rm{B}}}} =& {N_{tx,k}}{c_L}{q_1} \\
&+ \sum\limits_{i = 1}^{k - 1} \left[ \epsilon {N_{tx,i}}{c_L}{q_{k-i + 1}}  + \left( {1 - \epsilon } \right){N_{tx,i}}{c_H}{q_{k-i + 1}} \right],
\end{split}    
\end{align}
and
\begin{align}
\begin{split}
\label{sigma_H0_B}
\sigma _{{\mathcal{H}_0},k,{\rm{B}}}^2 = {N_{tx,k}}{c_L}{q_1}\left( {1 - {q_1}} \right) + \sum\limits_{i = 1}^{k - 1} {\sigma _{{\rm{ISI}},i,{\rm{B}}}^2}  + \sigma _{n,k,{\rm{B}}}^2,   
\end{split}    
\end{align}
where
\begin{align}
\begin{split}
\label{sigma_ISI_B}
\sigma _{{\rm{ISI}},i,{\rm{B}}}^2 
 =& \epsilon \left( {1 - \epsilon } \right){\left[ {{N_{tx, i}}{c_L}{q_{k-i + 1}}} \right]^2} \\
 &+ \epsilon \left( {1 - \epsilon } \right){\left[ {{N_{tx, i}}{c_H}{q_{k-i + 1}}} \right]^2} \\
 &+ \epsilon {N_{tx,i}}{c_L}{q_{k-i + 1}}\left( {1 - {q_{k-i + 1}}} \right) \\
 &+ \left( {1 - \epsilon } \right){N_{tx,i}}{c_H}{q_{k-i + 1}}\left( {1 - {q_{k-i + 1}}} \right)\\
 &- 2\epsilon \left( {1 - \epsilon } \right){N_{tx, i}}{c_L}{q_{k-i + 1}}{N_{tx,i}}{c_H}{q_{k-i + 1}},
\end{split}    
\end{align}

While under hypothesis $\mathcal{H}_1$, the mean number of received $\rm{A}$ and $\rm{B}$ molecules can be expressed as
\begin{align}
\begin{split}
\label{received_H1_A}
{N_{rx,k,{\rm{A}}}} &= {N_{tx,k}}\left( {1 - {c_H}} \right){q_1}+ \sum\limits_{i = 1}^{k - 1} {{N_{tx,i}}{c_{tx,i,{\rm{A}}}}{q_{k-i + 1}} + {N_{n,k,{\rm{A}}}}}, \\
\end{split}    
\end{align}
and
\begin{align}
\begin{split}
{N_{rx,k,{\rm{B}}}} &= {N_{tx,k}}{c_H}{q_1} + \sum\limits_{i = 1}^{k - 1} {{N_{tx,i}}{c_{tx,i,{\rm{B}}}}{q_{k-i + 1}} + {N_{n,k,{\rm{B}}}}}. 
\end{split}    
\end{align}
Therefore, the number of received $\rm{A}$ and $\rm{B}$ molecules under $\mathcal{H}_1$ follows can be approximated by the normal distribution,
${N_{rx,k,{\rm{A}}}} \sim \mathcal{N}\left( {{\mu _{{\mathcal{H}_1},{\rm{A}}}},\sigma _{{\mathcal{H}_1},{\rm{A}}}^2} \right)$, ${N_{rx,k,{\rm{B}}}} \sim \mathcal{N}\left( {{\mu _{{\mathcal{H}_1},{\rm{B}}}},\sigma _{{\mathcal{H}_1},{\rm{B}}}^2} \right)$. The mean and variance of received $\rm{A}$ molecules under $\mathcal{H}_1$ during the $k$th bit interval can be expressed as
\begin{align}
\begin{split}
{\mu _{{\mathcal{H}_1},k,{\rm{A}}}} =& {N_{tx,k}}\left( {1 - {c_H}} \right){q_1} + \sum\limits_{i = 1}^{k - 1} \left[ \epsilon {N_{tx,i}}\left( {1 - {c_L}} \right){q_{k-i + 1}} \right. \\
&\left. + \left( {1 - \epsilon } \right){N_{tx,i}}\left( {1 - {c_H}} \right){q_{k-i + 1}} \right],  
\end{split}    
\end{align}
and
\begin{align}
\begin{split}
\sigma _{{\mathcal{H}_1},k,{\rm{A}}}^2 = {N_{tx,k}}\left( {1 - {c_H}} \right){q_1}\left( {1 - {q_1}} \right) + \sum\limits_{i = 1}^{k - 1} {\sigma _{{\rm{ISI}},i,{\rm{A}}}^2}  + \sigma _{n,k,{\rm{A}}}^2,   
\end{split}    
\end{align}
where ${\sigma _{{\rm{ISI}},i,{\rm{A}}}^2}$ is given by (\ref{sigma_ISI_A}).
The mean and variance of received $\rm{B}$ molecules under $\mathcal{H}_1$ can be expressed as
\begin{align}
\begin{split}
{\mu _{{\mathcal{H}_1},k,{\rm{B}}}} =& {N_{tx,k}}{c_H}{q_1} \\
&+ \sum\limits_{i = 1}^{k - 1} \left[ \epsilon {N_{tx,i}}{c_L}{q_{k-i + 1}} + \left( {1 - \epsilon } \right){N_{tx,i}}{c_H}{q_{k-i + 1}} \right], 
\end{split}    
\end{align}
and
\begin{align}
\begin{split}
\label{sigma_H1_B}
\sigma _{{\mathcal{H}_1},k,{\rm{B}}}^2 = {N_{tx,k}}{c_H}{q_1}\left( {1 - {q_1}} \right) + \sum\limits_{i = 1}^{k - 1} {\sigma _{{\rm{ISI}},i,{\rm{B}}}^2}  + \sigma _{n,k,{\rm{B}}}^2,   
\end{split}    
\end{align}
where $\sigma _{{\rm{ISI}},i,{\rm{B}}}^2$ is given by (\ref{sigma_ISI_B}).


%
%

{\color{black}Since the transmitted molecules comprise a mixture of type $\rm{A}$ and type $\rm{B}$ molecules, and the ratio of A to B differs between the low and high reservoirs. Therefore, at the receiver, based on the Neyman-Pearson criterion, the maximum likelihood decision rule in the $k$th bit interval can be expressed as
\begin{align}
\begin{split}
\frac{{f_{\frac{{{N_{rx}},{\rm{A}}}}{{{N_{rx}},{\rm{B}}}}}^0}}{{f_{\frac{{{N_{rx}},{\rm{A}}}}{{{N_{rx}},{\rm{B}}}}}^1}} \underset{H_1}{\stackrel{H_0}{\gtrless}} \gamma 
\end{split},
\end{align}
where $f_{\frac{{{N_{rx}},{\rm{A}}}}{{{N_{rx}},{\rm{B}}}}}^0$ denotes the ratio of received A and B molecules under ${H_0}$, $f_{\frac{{{N_{rx}},{\rm{A}}}}{{{N_{rx}},{\rm{B}}}}}^1$ denotes the ratio of received A and B molecules under ${H_1}$, and $\gamma$ is the decision threshold. 
Therefore, at the receiver, we can employ ${{{N_{\text{rx},k,\text{A}}}}}/{{{N_{\text{rx},k,\text{B}}}}}$ to determine the received bits, and the decision rule can be expressed as
\begin{align}
{b_{rx,k}} = \left\{ {\begin{array}{*{20}{c}}
{0,}&{\frac{{{N_{rx,k,{\rm{A}}}}}}{{{N_{rx,k,{\rm{B}}}}}} > \gamma ,}\\
{1,}&{{\rm{otherwise.}}}
\end{array}} \right.   
\end{align}
}

Based on (\ref{received_A}) $\sim$ (\ref{sigma_ISI_B}), under ${\mathcal{H}_0}$, 
${N_{rx,k,{\rm{A}}}} - \gamma {N_{rx,k,{\rm{B}}}} \sim \mathcal{N}\left( {\mu_{\mathcal{H}_0,k},{\sigma_{\mathcal{H}_0,k}^2}} \right)$,
where $\mu_{\mathcal{H}_0,k}={{\mu _{{\mathcal{H}_0},k,{\rm{A}}}} - \gamma {\mu _{{\mathcal{H}_0},k,{\rm{B}}}}}$ and $\sigma_{\mathcal{H}_0,k}^2=\sigma _{{\mathcal{H}_0},k,{\rm{A}}}^2 + {\gamma ^2}\sigma _{{\mathcal{H}_0},k,{\rm{B}}}^2$. Based on (\ref{received_H1_A}) $\sim$ (\ref{sigma_H1_B}), under ${\mathcal{H}_1}$, ${N_{rx,k,{\rm{A}}}} - \gamma {N_{rx,k,{\rm{B}}}} \sim \mathcal{N}\left( \mu_{\mathcal{H}_1},\sigma_{\mathcal{H}_1}^2 \right)$, where $\mu_{\mathcal{H}_1,k}={\mu _{{\mathcal{H}_1},k,{\rm{A}}}} - \gamma {\mu _{{\mathcal{H}_1},k,{\rm{B}}}}$ and $\sigma_{\mathcal{H}_1,k}^2=\sigma _{{\mathcal{H}_1},k,{\rm{A}}}^2 + {\gamma ^2}\sigma _{{\mathcal{H}_1},k,{\rm{B}}}^2$. Therefore, 
$f\left( {{\rm{A}}|{{\cal H}_0},k} \right) - \gamma f\left( {{\rm{B}}|{{\cal H}_0},k} \right) = \frac{1}{{\sqrt {2\pi \sigma _{{{\cal H}_0},k}^2} }}\exp \left( { - \frac{{{{\left( {{N_{rx,k}} - {\mu _{{{\cal H}_0},k}}} \right)}^2}}}{{2\sigma _{{{\cal H}_0},k}^2}}} \right)$ and 
$f\left( {{\rm{A}}|{{\cal H}_1},k} \right) - \gamma f\left( {{\rm{B}}|{{\cal H}_1},k} \right) = \frac{1}{{\sqrt {2\pi \sigma _{{{\cal H}_1},k}^2} }}\exp \left( { - \frac{{{{\left( {{N_{rx,k}} - {\mu _{{{\cal H}_1},k}}} \right)}^2}}}{{2\sigma _{{{\cal H}_1},k}^2}}} \right)$. 

{\color{black} Obviously, the proposed detector effectively reduces the influence of interference molecules, consequently mitigating ISI. Furthermore, as the energy cost increases, the gap of $\rm{A}/\rm{B}$ molecules between the low and high reservoir widens. This reduction in released interference molecules leads to a more pronounced mitigation of ISI.}
The average probability of error ($P_e$)  from time slot 1 to $k$
can be expressed as \cite{varshney2018flow}
\begin{align}
{P_e} = \frac{1}{k}\sum\limits_{i = 1}^k {\left[ {\frac{1}{2}\left( {1 - Q\left( {\frac{{{\gamma} - {\mu _{{\mathcal{H}_1},i}}}}{{{\sigma _{{\mathcal{H}_1},i}}}}} \right)} \right) + \frac{1}{2}Q\left( {\frac{{{\gamma} - {\mu _{{\mathcal{H}_0},i}}}}{{{\sigma _{{\mathcal{H}_0},i}}}}} \right)} \right]}\label{p_e_lamda}    
\end{align}
The detailed calculation of $P_e$ is shown in the appendix.

\section{performance evaluation}
In this section, we analyze the performance of the considered MC system with an imperfect transmitter. In the simulation, the parameters are set in Table I. {\color{black} 
Especially, the Particle-Based Simulator (PBS) for MC is a simulation framework designed to emulate the behavior of individual molecules. In this simulation, the bit interval is discretized into time steps of duration $\Delta t$. Each step involves the movement of molecules, and their movement in each direction follows a Gaussian distribution with a mean of 0 and a standard deviation of $\sqrt{2D_m\Delta t}$, where $D_m$ is the diffusion coefficient of information molecules. Notably, the Signal-to-Noise Ratio (SNR), representing the ratio of the average squared number of observed molecules to the noise power\cite{luo2018one}, is set at 15dB.}
\begin{table}
\normalsize
\caption{SIMULATION PARAMETERS}
\centering
\begin{center}
\setlength{\tabcolsep}{1.5mm}
\begin{tabular}{  c  c  c  p{3cm}}
\hline
Symbol & Explanation & Value  \\ \hline
$D_m$ & Diffusion coefficient & $10^{-9}$ m$^2$/s \\ \hline
$d$ & Distance between transmitter and receiver & 10 $\mu$m  \\ \hline
$r$ & Radius of the receiver & 4 $\mu$m \\ \hline
$T$ & Bit interval & 1 s \\ \hline
$\Delta t$ & Discrete steps & 100 $\mu s$ \\ \hline
$N_m$ & Number of transmitted molecules &  1000 \\ \hline
$k$ & Boltzmann's constant & 1.3807 $\times$ $10^{-23}$ \\ \hline
$T_e$ & Absolute temperature & 298.15 \\ \hline
\end{tabular}
\end{center}
\end{table}

\begin{figure}[!t]
   \centering
   \includegraphics[width=0.47\textwidth]{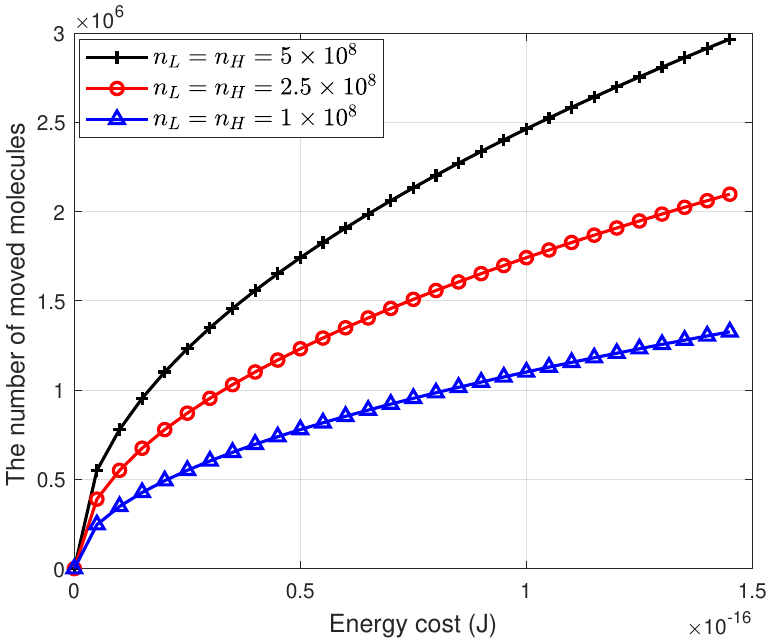}\\
   \caption{The number of moved molecules versus energy cost under $c = 0.5$.}\label{moved_molecules_n} 
\end{figure}

\begin{figure}[!t]
   \centering
   \includegraphics[width=0.47\textwidth]{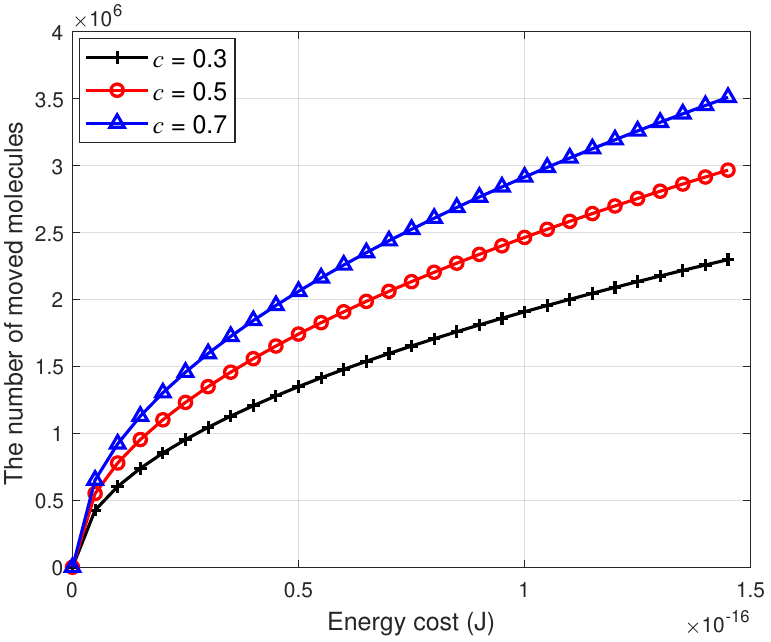}\\
   \caption{The number of moved molecules versus energy cost under $n_L = n_H =5 \times 10^8$.}\label{moved_molecules_c} 
\end{figure}

In Figs. \ref{moved_molecules_n} and \ref{moved_molecules_c}, the number of molecules moved between reservoirs is shown to vary with the energy cost{\color{black}, measured in Joule (J)}, for different numbers of molecules in the reservoirs, {\color{black} as derived in (7)}. The number of moved molecules increases with the energy cost, and for the given energy cost, for more $\rm{B}$ molecules (as shown in Fig. \ref{moved_molecules_n}) or higher concentration of $\rm{B}$ molecules in the reservoirs (as shown in Fig. \ref{moved_molecules_c}), a larger number of molecules can be moved. With the increase of moved molecules, the lower concentration of molecules in the reservoir, then, the larger energy cost is required. These results illustrate that creating a near-perfect transmitter (where each reservoir contains only one kind of molecule) is very costly.

\begin{figure}[!t]
   \centering
   \includegraphics[width=0.47\textwidth]{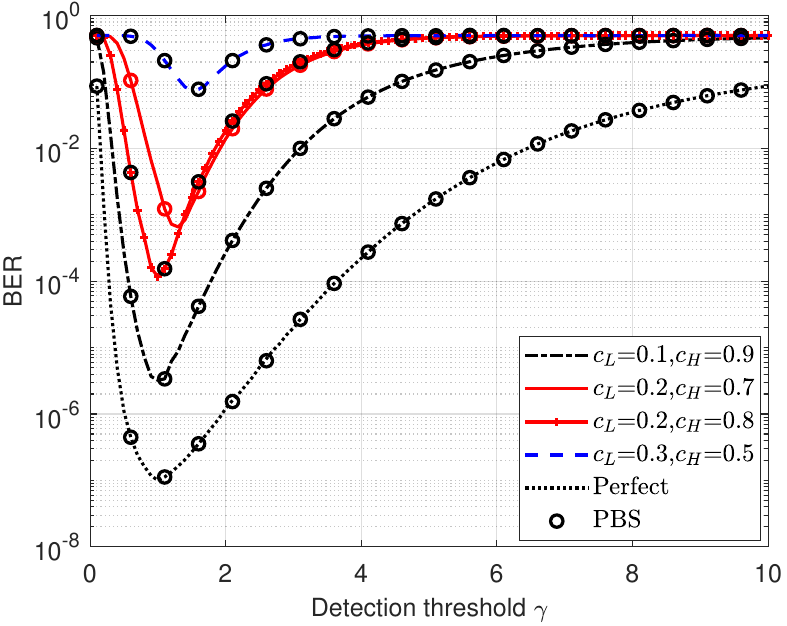}\\
   \caption{The BER versus the detection threshold $\gamma$.}\label{ber_threshold} 
\end{figure}

In Fig. \ref{ber_threshold}, we analyze how the BER varies with detection threshold $\gamma$ under different mole fractions of $\rm{B}$ molecules in the low and high reservoirs, {\color{black} as derived in (\ref{p_e_lamda})}. It is shown in Fig. \ref{ber_threshold} that, for a given mole fraction of molecules in the transmitter, there exists an optimal detection threshold, which varies with the mole fractions of the transmitter. The system achieves better performance for the larger difference of the $\rm{B}$ mole fraction between the low and high reservoirs.
In the figure, the ``perfect'' line indicates the perfect transmitter which means that there are only $\rm{A}$ molecules in the low reservoir and only $\rm{B}$ molecules in the high reservoir, as there is no interference molecule from the transmitter; therefore, it achieves the best performance. Thus, the figure indicates that the system performance can be improved at the cost of energy consumption. As shown in Fig. 8, compared to $c_L=0.2$ and $c_H=0.7$, $c_L=0.2$ and $c_H=0.8$ achieves lower BER, as the ratio of $c_L=0.2$ and $c_H=0.8$ is larger than $c_L=0.2$ and $c_H=0.7$,  which results in a more substantial gap in the number of received molecules between A and B. Consequently, when employing the proposed detection scheme, $c_L=0.2$ and $c_H=0.8$ achieves better BER performance. Moreover, PBS is conducted to validate the analytical results.

\begin{figure}[!t]
   \centering
   \includegraphics[width=0.47\textwidth]{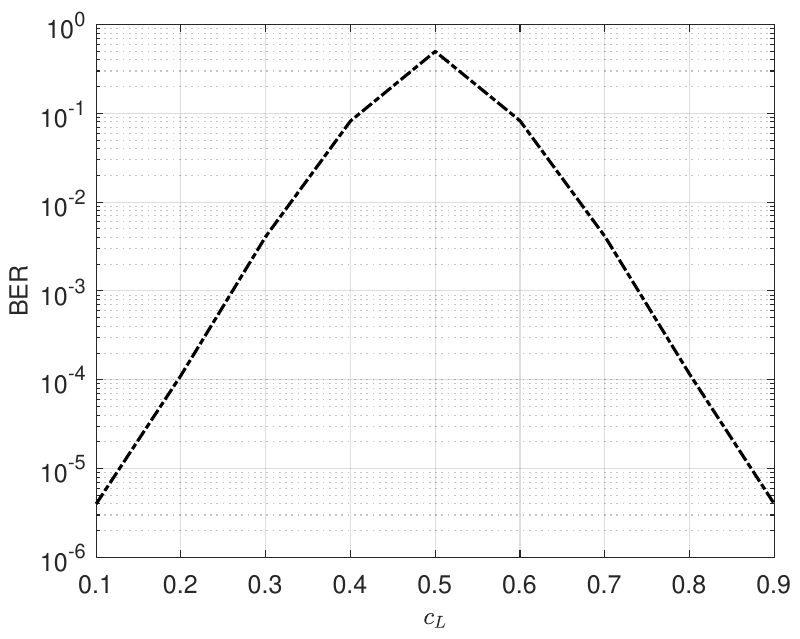}\\
   \caption{The BER versus the initial fraction of $\rm{B}$ molecules in the low reservoir $c_L$.}\label{BER_cL_v2}
\end{figure}

Fig. \ref{BER_cL_v2} shows how the BER varies with $c_L$, which is defined as the mole fraction of $\rm{B}$ molecules in the low reservoir, {\color{black}as derived from (25)-(36) and (39).} 
Recalling the transmission scheme for transmitting each symbol, it can be seen from Fig. \ref{BER_cL_v2}, when $c_L$ is farther from 0.5, the performance of the system is better; this is because when $c_L$ is farther from 0.5, the fraction of signal molecules is much larger than the interfering molecules. This result also reflects the result in Fig. \ref{ber_threshold}.

\begin{figure}[!t]
   \centering
   \includegraphics[width=0.47\textwidth]{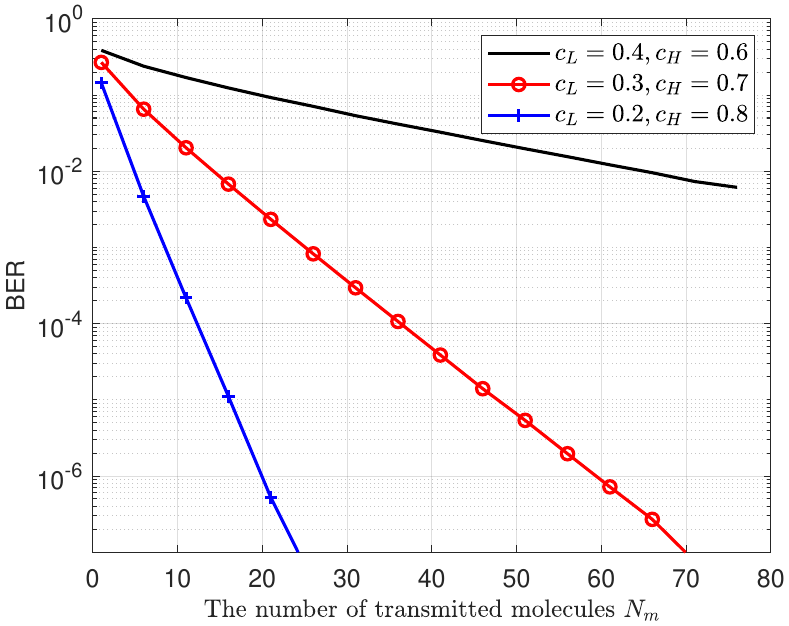}\\
   \caption{The BER versus the number of transmitted molecules $N_m$ without ISI and counting noise.}\label{BER_Nm_v3}
\end{figure}

\begin{figure}[!t]
   \centering
   \includegraphics[width=0.47\textwidth]{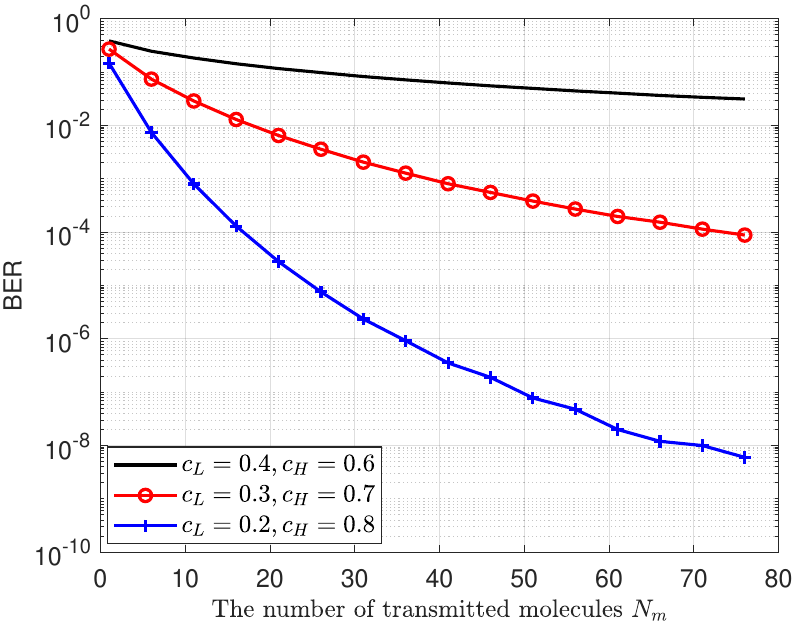}\\
   \caption{The BER versus the number of transmitted molecules $N_m$ without ISI but considering the counting noise.}\label{BER_Nm_noise_v4}
\end{figure}

Figs. \ref{BER_Nm_v3} and \ref{BER_Nm_noise_v4} demonstrate how the BER varies with the number of transmitted molecules $N_m$, and clearly show that the BER decreases with the increase of $N_m$. In Fig. \ref{BER_Nm_v3}, neither ISI nor counting noise are considered, while in Fig. \ref{BER_Nm_noise_v4}, only ISI is considered. 
Thus, in Fig. \ref{BER_Nm_v3}, BER arises only from the imperfections in the transmitter. Therefore, as the number of transmitted molecules increases, the error probability of selected molecules from the reservoirs decreases. Furthermore, with a larger gap between the $c_L$ and $c_H$, the concentration of $\rm{A}$ molecules in the low reservoir and the concentration of $\rm{B}$ molecules in the high reservoir are both larger; thus, the BER decreases as the error probability of molecules transmitted from the reservoirs is lower. Meanwhile, in Fig. \ref{BER_Nm_noise_v4}, as counting noise is considered which is related to the number of received molecules and the volume of the receiver, therefore, 
the slope of the BER decreases with the increase of $N_m$.

\begin{figure}[!t]
   \centering
   \includegraphics[width=0.47\textwidth]{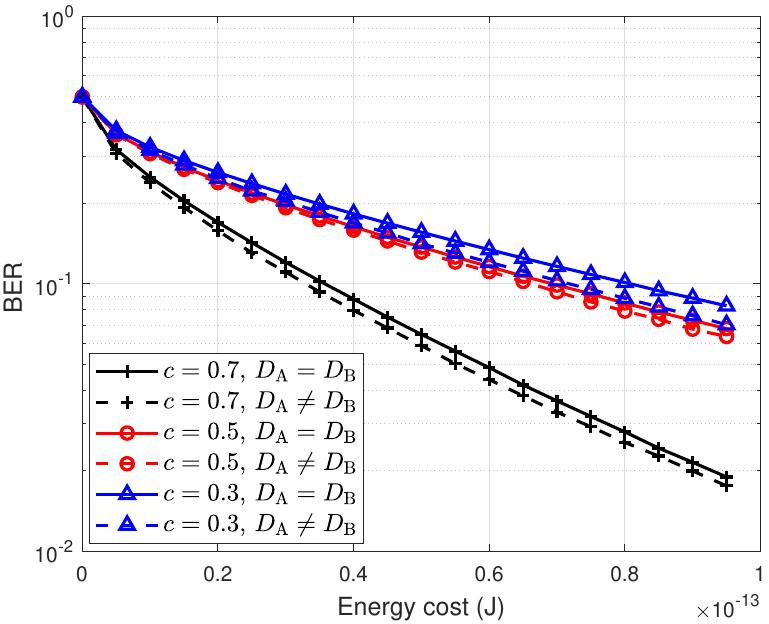}\\
   \caption{The BER versus energy cost under the different initial concentrations of $\rm{B}$. For $D_{\rm{A}}=D_{\rm{B}}=10^{-9}$, for $D_{\rm{A}}\ne D_{\rm{B}}$, $D_{\rm{A}}=10^{-9}$, $D_{\rm{B}}=10^{-8}$.}\label{BER_energy}
\end{figure}

In Fig. \ref{BER_energy}, we show how the BER varies with the energy cost under differing initial concentrations of $\rm{B}$ molecules in the reservoirs both considering ISI and counting noise {\color{black}as derived from (25)-(36) and (39)}. As shown in Fig. \ref{BER_energy}, the BER decreases with the increase of energy cost, as with the increase of energy cost, more $\rm{B}$ molecules are moved from the low reservoir to the high reservoir, then the gap of $\rm{A}/\rm{B}$ molecules between the low and high reservoir is larger, therefore, the BER decreases. {\color{black}And due to a reduction in released interference molecules, the ISI can also be mitigated, albeit at the expense of an increased energy cost.} For the larger $c$, under the given energy cost, more $\rm{B}$ molecules are moved, and the gap is also larger compared to a small $c$, therefore, it achieves better BER performance. Fig. \ref{BER_energy} also indicates that the slope of the BER decreases with the increase of energy cost, indicating a problem of diminishing returns.  As shown in Fig. 8, for different diffusion coefficients, the system achieves better BER performance, which is due to one diffusion coefficient being significantly larger: in this case, the released molecules more quickly diffuse away, resulting in less ISI, and better BER performance.
%
%

\begin{figure}[!t]
   \centering
   \includegraphics[width=0.47\textwidth]{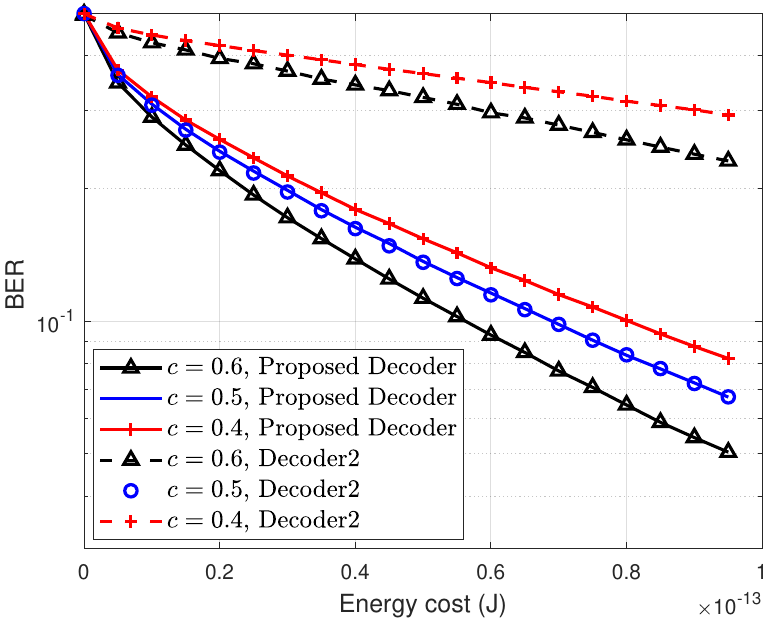}\\
   \caption{The BER performance is compared between the proposed detection method and the traditional detection method, referred to as Decoder2 in the figure.}\label{ber_energy_imperfet_ISI_v2}
\end{figure}

{\color{black}In Fig. \ref{ber_energy_imperfet_ISI_v2}, we compare the BER performance between the proposed detection method, which is derived in (37), and the conventional detection method (Decoder2), which compares the number of received different types of molecules. As shown in Fig. \ref{ber_energy_imperfet_ISI_v2}, when $c=0.5$, the two decoders achieve the same performance. This is because, when $c=0.5$, i.e. there are the same number of A and B molecules initially in the reservoirs, then after moving B molecules from the low reservoir to the high reservoir, there are more A molecules in the low reservoir and more B molecules in the high reservoir; therefore, the same performance is achieved, as seen in the figure. However, when $c\ne0.5$, after moving B molecules from the low reservoir to the high reservoir, it may no longer be the case that there are more A molecules in the low reservoir and more B molecules in the high reservoir, e.g. when $c=0.6$, for constant energy cost, after moving B molecules, maybe there are still more B molecules both in the low and high reservoir, making the BER performance of Decoder2 decrease. Therefore, the proposed detector can effectively reduce the impact of interference molecules, simultaneously mitigating ISI.}

\section{conclusion}
In this paper, a molecular communication system with an imperfect transmitter is considered, in which the transmitter contains reservoirs with different concentrations of message-bearing molecules. Unlike the conventional assumption of a perfect transmitter, the molecules in the reservoirs are mixed, but there is a difference between the concentration of different types of molecules. Moreover, our system explicitly takes into account the chemical potential energy required to create the transmitter. Under these assumptions, a detection method based on the ratio of different types of molecules is proposed, and
the average BER is derived. Simulation results showed that, with the increase of the difference of concentrations, the system achieves better BER performance, and the ideal transmitter achieves the best performance. However, there is a tradeoff between increased performance and energy expenditure, and we are able to quantify that tradeoff. The imperfect transmitter can also be applied to CSK modulation with a single reservoir, potentially offering a simpler alternative by adjusting the number of molecules that are emitted and mixed with those molecules outside of the reservoir, which would be an interesting topic of future research.

\appendix 
\section{}
Here we give a detailed calculation of $P_e$ in $\left(\ref{p_e_lamda}\right)$. This can be detailed expressed as
\begin{align}
\begin{split}
{P_e} &= \frac{1}{k}\sum\limits_{i = 1}^k {\left[ {\frac{1}{2}\left( {1 - Q\left( {\frac{{\gamma  - {\mu _{{{\cal{H}}_1},i}}}}{{{\sigma _{{{\cal{H}}_1},i}}}}} \right)} \right) + \frac{1}{2}Q\left( {\frac{{\gamma  - {\mu _{{{\cal{H}}_0},i}}}}{{{\sigma _{{{\cal{H}}_0},i}}}}} \right)} \right]} \\
 &= \frac{1}{k}\sum\limits_{i = 1}^k \left[ \frac{1}{2}\left( {1 - Q\left( {\frac{\gamma  - \left( {\mu _{{{\cal{H}}_1},\rm{A},k}} - \gamma {\mu _{{{\cal{H}}_1},\rm{B},k}} \right)}{\sqrt {\sigma _{{{\cal{H}}_1},{\rm{A}},k}^2 + {\gamma ^2}\sigma _{{{\cal{H}}_1},\rm{B},k}^2} }} \right)} \right) \right. \\
 &\left. + \frac{1}{2}Q\left( {\frac{{\gamma  - \left( {{\mu _{{{\cal{H}}_0},\rm{A},k}} - \gamma {\mu _{{{\cal{H}}_0},\rm{B},k}}} \right)}}{{\sqrt {\sigma _{{{\cal{H}}_0},{\rm{A}},k}^2 + {\gamma ^2}\sigma _{{{\cal{H}}_0},\rm{B},k}^2} }}} \right) \right].
\end{split}
\end{align}

Assuming $n_L=n_H = \frac{1}{2}n$, then ${\mu _{{{\cal{H}}_0},\rm{A},k}}$ can be expressed as
\begin{align}
\begin{split}
{\mu _{{{\cal{H}}_0},\rm{A},k}}  &= {N_{tx,k}}\left( {1 - {c_L}} \right){q_1} + \sum\limits_{i = 1}^{k - 1} \left[ \epsilon {N_{tx,i}}\left( {1 - {c_L}} \right){q_{k-i + 1}} \right.\\
&\left. + \left( {1 - \epsilon } \right){N_{tx,i}}\left( {1 - {c_H}} \right){q_{k-i + 1}} \right], \\
&= {N_{tx,k}}\left( {1 - {c_L}} \right){q_1} + \sum\limits_{i = 1}^{k - 1}  \left[\frac{1}{2}{N_{tx,i}}\left( {1 - {c_L}} \right)q_{k - i + 1} \right. \\
&\left. + \frac{1}{2}{N_{tx,i}}\left( {1 - {c_H}} \right){q_{k - i + 1}}\right],\\  
& = {N_{tx,k}}\left( {1 - c - \sqrt {\frac{{cn}}{{2kT{n_L}}}E} } \right){q_1} \\
&+ \sum\limits_{i = 1}^{k - 1} \left[ \frac{1}{2}{N_{tx,i}}\left( {1 - c - \sqrt {\frac{{cn}}{{2kT{n_L}}}E} } \right){q_{k - i + 1}}\right.\\
&\left. + \frac{1}{2}{N_{tx,i}}\left( {1 - c + \sqrt {\frac{{cn}}{{2kT{n_H}}}E} } \right){q_{k - i + 1}} \right]\\
&= {N_{tx,k}}\left( {1 - c - \sqrt {\frac{c}{{kT}}E} } \right){q_1} \\
&+ \sum\limits_{i = 1}^{k - 1} \left[ {N_{tx,i}}\left( {1 - c} \right){q_{k - i + 1}} \right].
\end{split}
\end{align}
The quantity ${\mu _{{{\cal H}_0},{\rm{B}},k}}$ can be expressed as
\begin{align}
\begin{split}
{\mu _{{{\cal H}_0},{\rm{B}},k}} &= {N_{tx,k}}{c_L}{q_1} \\
&+\sum\limits_{i = 1}^{k - 1} {\left[ {{\epsilon N_{tx,i}}{c_L}{q_{k - i + 1}} + \left( {1 - \epsilon } \right){N_{tx,i}}{c_H}{q_{k - i + 1}}} \right]} \\
& = {N_{tx,k}}{c_L}{q_1} + \sum\limits_{i = 1}^{k - 1} \left[ {\frac{1}{2} N_{tx,i}}{c_L}{q_{k - i + 1}}  +  {\frac{1}{2} } {N_{tx,i}}{c_H}{q_{k - i + 1}} \right]  \\
& = {N_{tx,k}}\left( {c - \sqrt {\frac{{cn}}{{2kT{n_L}}}E} } \right){q_1} \\
&+ \sum\limits_{i = 1}^{k - 1} \left[ \frac{1}{2}{N_{tx,i}}\left( {c - \sqrt {\frac{{cn}}{{2kT{n_L}}}E} } \right){q_{k - i + 1}} \right.\\
& \left. + \frac{1}{2}{N_{tx,i}}\left( {c + \sqrt {\frac{{cn}}{{2kT{n_H}}}E} } \right){q_{k - i + 1}} \right] \\
& = {N_{tx,k}}\left( {c - \sqrt {\frac{c}{{kT}}E} } \right){q_1} + \sum\limits_{i = 1}^{k - 1} {N_{tx,i}}c{q_{k - i + 1}}.  
\end{split}    
\end{align}

The quantity ${\mu _{{{\cal H}_1},{\rm{A}},k}}$ can be expressed as
\begin{align}
\begin{split}
{\mu _{{{\cal H}_1},{\rm{A}},k}} &= {N_{tx,k}}\left( {1 - {c_H}} \right){q_1} + \sum\limits_{i = 1}^{k - 1} \left[\epsilon {N_{tx,i}}\left( {1 - {c_L}} \right){q_{k - i + 1}} \right. \\ & \left. + \left( {1 - \epsilon} \right){N_{tx,i}}\left( {1 - {c_H}} \right){q_{k - i + 1}} \right]  \\
&= {N_{tx,k}}\left( {1 - c + \sqrt {\frac{c}{{kT}}E} } \right){q_1} \\
&+ \sum\limits_{i = 1}^{k - 1} \left[ \frac{1}{2}{N_{tx,i}}\left( {1 - \left( {c - \sqrt {\frac{c}{{kT}}E} } \right)} \right){q_{k - i + 1}} \right. \\  & \left. + \frac{1}{2}{N_{tx,i}}\left( {1 - \left( {c + \sqrt {\frac{c}{{kT}}E} } \right)} \right){q_{k - i + 1}} \right]  \\
&= {N_{tx,k}}\left( {1 - c + \sqrt {\frac{c}{{kT}}E} } \right){q_1}\\ 
&+ \sum\limits_{i = 1}^{k - 1} {{N_{tx,i}}\left( {1 - c} \right){q_{k - i + 1}}} .
\end{split}
\end{align}

The quantity ${\mu _{{{\cal H}_1},{\rm{B}},k}}$ can be expressed as
\begin{align}
\begin{split}
{\mu _{{{\cal H}_1},{\rm{B}},k}} &= {N_{tx,k}}{c_H}{q_1}\\
&+ \sum\limits_{i = 1}^{k - 1} \left[ \epsilon {N_{tx,i}}{c_L}{q_{k - i + 1}}+ \left( {1 - \epsilon} \right){N_{tx,i}}{c_H}{q_{k - i + 1}} \right]  \\
&= {N_{tx,k}}\left( {c + \sqrt {\frac{{cn}}{{2kT{n_H}}}E} } \right){q_1} \\
&+ \sum\limits_{i = 1}^{k - 1} \left[ \frac{1}{2}{N_{tx,i}}\left( {c - \sqrt  {\frac{{cn}}{{2kT{n_L}}}E} } \right){q_{k - i + 1}} \right.\\ 
&\left. + \frac{1}{2}{N_{tx,i}}\left( {c + \sqrt {\frac{{cn}}{{2kT{n_H}}}E} } \right){q_{k - i + 1}} \right]  \\
&= {N_{tx,k}}\left( {c + \sqrt {\frac{c}{{kT}}E} } \right){q_1} + \sum\limits_{i = 1}^{k - 1} {{N_{tx,i}}c{q_{k - i + 1}}} .
\end{split}
\end{align}

The quantities $\sigma _{{\mathcal{H}_0},{\rm{A}},k}^2$ and $\sigma _{{\mathcal{H}_1},{\rm{A}},k}^2$ can be expressed as
\begin{align}
\begin{split}
\sigma _{{\mathcal{H}_0},{\rm{A}},k}^2 = {N_{tx,k}}\left( {1 - {c_L}} \right){q_1}\left( {1 - {q_1}} \right) + \sum\limits_{i = 1}^{k - 1} {\sigma _{{\rm{ISI}},{\rm{A}},i}^2}  + \sigma _{n,{\rm{A}},k}^2,   
\end{split}    
\end{align}

\begin{align}
\begin{split}
\sigma _{{\mathcal{H}_1},{\rm{A}},k}^2 = {N_{tx,k}}\left( {1 - {c_H}} \right){q_1}\left( {1 - {q_1}} \right) + \sum\limits_{i = 1}^{k - 1} {\sigma _{{\rm{ISI}},{\rm{A}},i}^2}  + \sigma _{n,{\rm{A}},k}^2,   
\end{split}    
\end{align}

\begin{align}
\begin{split}
\sigma _{{\mathcal{H}_0},{\rm{B}},k}^2 = {N_{tx,k}}{c_L}{q_1}\left( {1 - {q_1}} \right) + \sum\limits_{i = 1}^{k - 1} {\sigma _{{\rm{ISI}},{\rm{B}},i}^2}  + \sigma _{n,{\rm{B}},k}^2,   
\end{split}    
\end{align}

\begin{align}
\begin{split}
\sigma _{{\mathcal{H}_1},{\rm{B}},k}^2 = {N_{tx,k}}{c_H}{q_1}\left( {1 - {q_1}} \right) + \sum\limits_{i = 1}^{k - 1} {\sigma _{{\rm{ISI}},{\rm{B}},i}^2}  + \sigma _{n,{\rm{B}},k}^2,   
\end{split}    
\end{align}

where
\begin{align}
\begin{split}
\sigma _{{\rm{ISI}},{\rm{B}},i}^2 &= \epsilon \left( {1 - \epsilon } \right){\left[ {{N_{tx,i}}{c_L}{q_{k - i + 1}}} \right]^2}+ \epsilon \left( {1 - \epsilon} \right){\left[ {{N_{tx,i}}{c_H}{q_{k - i + 1}}} \right]^2} \\
&+ \epsilon {N_{tx,i}}{c_L}{q_{k - i + 1}}\left( {1 - {q_{k - i + 1}}} \right) \\
&+ \left( {1 - \epsilon} \right){N_{tx,i}}{c_H}{q_{k - i + 1}}\left( {1 - {q_{k - i + 1}}} \right)\\ 
&- 2\epsilon \left( {1 - \epsilon} \right){N_{tx,i}}{c_L}{q_{k - i + 1}}{N_{tx,i}}{c_H}{q_{k - i + 1}}\\
 &= \frac{1}{4}{\left[ {{N_{tx,i}}\left( {c - \sqrt {\frac{{cn}}{{2kT{n_L}}}E} } \right){q_{k - i + 1}}} \right]^2} \\
 &+ \frac{1}{4}{\left[ {{N_{tx,i}}\left( {c + \sqrt {\frac{{cn}}{{2kT{n_H}}}E} } \right){q_{k - i + 1}}} \right]^2} \\
 &+ \frac{1}{2}{N_{tx,i}}\left( {c - \sqrt {\frac{{cn}}{{2kT{n_L}}}E} } \right){q_{k - i + 1}}\left( {1 - {q_{k - i + 1}}} \right) \\ 
 &+ \frac{1}{2}{N_{tx,i}}\left( {c + \sqrt {\frac{{cn}}{{2kT{n_H}}}E} } \right){q_{k - i + 1}}\left( {1 - {q_{k - i + 1}}} \right) \\
 &- \frac{1}{2}{N_{tx,i}}\left( {c - \sqrt {\frac{{cn}}{{2kT{n_L}}}E} } \right){q_{k - i + 1}} \\
 &\times {N_{tx,i}}\left( {c + \sqrt {\frac{{cn}}{{2kT{n_H}}}E} } \right){q_{k - i + 1}}\\
 &= N_{tx,i}^2\frac{c}{{kT}}Eq_{k - i + 1}^2 + {N_{tx,i}}c{q_{k - i + 1}}\left( {1 - {q_{k - i + 1}}} \right).
\end{split}
\end{align}

\begin{align}
\begin{split}
\sigma _{{\rm{ISI}},{\rm{A}},i}^2 &= \epsilon \left( {1 - \epsilon} \right){\left[ {{N_{tx,i}}\left( {1 - {c_L}} \right){q_{k - i + 1}}} \right]^2} \\
&+ \epsilon \left( {1 - \epsilon} \right){\left[ {{N_{tx,i}}\left( {1 - {c_H}} \right){q_{k - i + 1}}} \right]^2} \\
&+ \epsilon {N_{tx,i}}\left( {1 - {c_L}} \right){q_{k - i + 1}}\left( {1 - {q_{k - i + 1}}} \right) \\
&+ \left( {1 - \epsilon} \right){N_{tx,i}}\left( {1 - {c_H}} \right){q_{k - i + 1}}\left( {1 - {q_{k - i + 1}}} \right) \\
&- 2\epsilon \left( {1 - \epsilon} \right){N_{tx,i}}\left( {1 - {c_L}} \right){q_{k - i + 1}}{N_{tx,i}}\left( {1 - {c_H}} \right){q_{k - i + 1}}\\
 &= \frac{1}{4} N_{tx,i}^2\left( 1 - 2\left( c - \sqrt {\frac{c}{kT}E} \right) \right.\\
 &\left. + \left( {c^2} - 2c\sqrt {\frac{c}{{kT}}E}  + \frac{c}{kT}E \right)\right)q_{k - i + 1}^2 \\
 &+ \frac{1}{4} N_{tx,i}^2\left( 1 - 2\left( {c + \sqrt {\frac{c}{{kT}}E} } \right) \right.\\
 &\left. + \left( {{c^2} + 2c\sqrt {\frac{c}{{kT}}E}  + \frac{c}{{kT}}E} \right) \right)q_{k - i + 1}^2 \\
 &+ \frac{1}{2}{N_{tx,i}}\left( {1 - \left( {c - \sqrt {\frac{c}{{kT}}E} } \right)} \right){q_{k - i + 1}}\left( {1 - {q_{k - i + 1}}} \right) \\
 &+ \frac{1}{2}{N_{tx,i}}\left( {1 - \left( {c + \sqrt {\frac{c}{{kT}}E} } \right)} \right){q_{k - i + 1}}\left( {1 - {q_{k - i + 1}}} \right) \\
 &- \frac{1}{2}N_{tx,i}^2\left( {1 - \left( {c - \sqrt {\frac{c}{{kT}}E} } \right)} \right)\\
 &\times\left( {1 - \left( {c + \sqrt {\frac{c}{{kT}}E} } \right)} \right)q_{k - i + 1}^2\\
 &= N_{tx,i}^2\frac{c}{{kT}}Eq_{k - i + 1}^2 + {N_{tx,i}}\left( {1 - c} \right){q_{k - i + 1}}\left( {1 - {q_{k - i + 1}}} \right).
\end{split}
\end{align}

The variance of the counting noise of type $\rm{A}$ molecules can be expressed as
\begin{align}
\begin{split}
\sigma _{n,{\rm{A}},k}^2 = {{N_{rx,{\rm{A}},k}}}/{{V_{rx}}},
\end{split}
\end{align}

The variance of the counting noise of type $\rm{B}$ molecules can be expressed as
\begin{align}
\begin{split}
\sigma _{n,{\rm{B}},k}^2 = {{N_{rx,{\rm{B}},k}}}/{{V_{rx}}}.
\end{split}
\end{align}

\bibliographystyle{IEEEtran}
\bibliography{references}
\end{document}